\documentclass[conference]{IEEEtran}
\usepackage{amsmath, amssymb, bm, cite, epsfig, psfrag}
\usepackage{graphicx}
\usepackage{soul} 
\usepackage[left=0.625in, right=0.625in, top=0.7in, bottom=0.95in]{geometry}
\usepackage{dblfloatfix}
\usepackage{array}
\usepackage{lipsum}
\usepackage{subcaption}

\newcolumntype{P}[1]{>{\centering\hspace{0pt}}p{#1}}
\newcolumntype{M}[1]{>{\centering\hspace{0pt}}m{#1}}
\newcolumntype{L}{>{\centering\arraybackslash}m{3cm}}
\usepackage{caption}
\usepackage{epstopdf}	
\usepackage{longtable}
\usepackage{supertabular,booktabs}
\usepackage{bbm}
\usepackage{multirow}
\usepackage[usenames,dvipsnames]{xcolor}

\usepackage{etoolbox}
\usepackage{pbox}
\usepackage{fixltx2e}
\usepackage{tabu}	
\usepackage{filecontents}
\usepackage{enumerate}
\usepackage{textcomp}
\usepackage{makecell}
\usepackage{gensymb}
\usepackage{colortbl}
\usepackage{fancyhdr}
\def\PL{\mathrm{PL}}
\def\dB{\mathrm{dB}}
\newtoggle{conference}

\togglefalse{conference} 
\graphicspath{{figures/}}

\newcolumntype{?}{!{\vrule width 2pt}}

\fancypagestyle{firststyle}{
	\fancyhf{}
	\fancyhead[L]{S. Ju and T. S. Rappaport, ``142 GHz Multipath Propagation Measurements and Path Loss Channel Modeling in Factory Buildings,''  2023 \textit{IEEE International Conference on Communications (ICC)}, May. 2023, pp. 1-6.}     %% C or L or R.
}
\IEEEoverridecommandlockouts

\begin{document}
\title{142 GHz Multipath Propagation Measurements and Path Loss Channel Modeling in Factory Buildings} 
\author{\IEEEauthorblockN{Shihao Ju and Theodore S. Rappaport}

\IEEEauthorblockA{	\small NYU WIRELESS, Tandon School of Engineering, New York University, Brooklyn, NY, 11201\\
				\{shao, tsr\}@nyu.edu}
					\thanks{This research is supported by the NYU WIRELESS Industrial Affiliates Program and National Science Foundation (NSF) Research Grants: 1909206 and 2037845.}
}

\maketitle
\thispagestyle{firststyle}
\begin{abstract}
This paper presents sub-Terahertz (THz) radio propagation measurements at 142 GHz conducted in four factories with various layouts and facilities to explore sub-THz wireless channels for smart factories in 6G and beyond. Here we study spatial and temporal channel responses at 82 transmitter-receiver (TX-RX) locations across four factories in the New York City area and over distances from 5 m to 85 m in both line-of-sight (LOS) and non-LOS (NLOS) environments. The measurements were performed with a sliding-correlation-based channel sounder with 1 GHz RF bandwidth with steerable directional horn antennas with 27 dBi gain and 8\degree~half-power beamwidth at both TX and RX, using both vertical and horizontal antenna polarizations, yielding over 75,000 directional power delay profiles. Channel measurements of two RX heights at 1.5 m (high) emulating handheld devices and at 0.5 m (low) emulating automated guided vehicles (AGVs) were conducted for automated industrial scenarios with various clutter densities. Results yield the first path loss models for indoor factory (InF) environments at 142 GHz and show the low RX height experiences a mean path loss increase of 10.7 dB and 6.0 dB when compared with the high RX height at LOS and NLOS locations, respectively. Furthermore, flat and rotatable metal plates were leveraged as passive reflecting surfaces (PRSs) in channel enhancement measurements to explore the potential power gain on sub-THz propagation channels, demonstrating a range from 0.5 to 22 dB improvement with a mean of 6.5 dB in omnidirectional channel gain as compared to when no PRSs are present.
	
\end{abstract}


\section{Introduction} \label{sec:intro}
Terahertz (THz) communications show great promise for sixth-generation (6G) wireless systems due to the vast available bandwidth \cite{Rap19Access,Rap13access}. Federal Communications Commission (FCC) released the first report and order, ET Docket 18-21, in 2019 \cite{FCC19ET18-21}, which offered experimental licenses for the 95 GHz to 3 THz range and opened up 21.2 GHz of the spectrum between 116 GHz and 246 GHz for unlicensed use. Extensive propagation measurements and accurate channel modeling are paramount to building a high-performance THz wireless system with communication and sensing functions. Significant efforts on channel measurements and modeling above 100 GHz have been made over the past few years \cite{Rap19Access,Ju21JSAC,Xing21CLb,Chen21TWC}, but there is virtually no understanding of real-world factory channels \cite{Ju22ICC}. Most measurement campaigns focused on indoor office and outdoor urban environments between 100 and 300 GHz, which have shown that, despite the fewer observed multipath components than mmWave frequencies, the non-line-of-sight (NLOS) propagation created by specular reflection and diffuse scattering is remarkably substantial in sub-THz wireless system design and deployment \cite{Ju21JSAC,Xing21CLb,Ju19ICC}. 

Advanced industrial Internet of things (IIOT) with increasing automated guided vehicles (AGVs) and autonomous mobile robots (AMRs) has rapidly transformed modern factories, creating an unprecedented demand for fast and reliable in-factory communication networks and centimeter-level positioning and sensing \cite{Chen18access}. Therefore, the ultrawide transmission bandwidth at sub-THz frequencies, together with the steerable pencil beams generated by large antenna arrays with a massive number of antenna elements, may satisfy the stringent requirements for communication and sensing in smart factories. 

This work presents a recent 142 GHz radio propagation measurement campaign in four factory buildings in Brooklyn, NY. Extensive channel measurements were conducted at 16 transmitter (TX) locations and 60 receiver (RX) locations across four factories, resulting in 30 line-of-sight (LOS) location pairs, 48 NLOS location pairs, and four outage location pairs over a T-R separation distance range from 5 m to 85 m. The TX antenna was vertically polarized, and the RX antenna was systematically switched between vertical (V) and horizontal (H) polarizations, providing co- and cross-polarized channel characterization. Directional and omnidirectional path loss models for co- and cross-polarization antenna configurations are derived in Section \ref{sec:pathloss} with a statistical cross-polarization discrimination (XPD) characterization. 

The measurement setup mimicked an indoor downlink transmission from an access point to user equipment with TXs at 2.5 m and RXs at 1.5 m heights above the floor in the four factories, labeled A, B, C, and D. At Factory C and D, additional channel measurements were conducted with the RX height of 0.5 m to investigate the channel properties of near-the-ground mobile receivers (e.g., AGVs) commonly used in factories. The measurement results in Section \ref{sec:rx_height} show that the low-height RXs experience larger path loss than the high-height RXs due to increasing blockage probability and complex scattering environment caused by surrounding clutter. 

The larger path loss at the low RX height motivates us to propose a channel enhancement scheme using passive reflecting surfaces (PRS). There have been some early works employing metal surfaces to enhance mmWave and sub-THz propagation channel by fixing a metal plate at the corner of an L-shaped corridor \cite{Khawaja20OJCS} or attaching aluminum foil to walls \cite{Li22arXiv} and pillars \cite{Abbasi21ICC}. Section \ref{sec:surface} demonstrates a new measurement setup using flat metal plates to improve propagation channels. A 1 m x 1 m flat smooth aluminum plate was placed near the walls of factories to increase the received power for a specified RX location by either boosting an existing weak multipath component (MPC) or creating an additional MPC. The PRS was set next to walls and pillars and manually rotated in the horizontal plane to the best reflecting orientation, which mimicked the flexible beamforming operation of future reconfigurable intelligent surfaces (RIS) at sub-THz frequencies \cite{Ellingson21PIMRC}. The PRS-aided channel measurements were conducted at all twelve locations at Factory D with both RX heights of 1.5 m and 0.5 m for comparative analysis. 

\section{142 GHz Propagation Measurements in Four Factories}
\label{sec:meas}
\subsection{Measurement Equipment}
The 142 GHz wideband channel sounder adopts a sliding-correlation-based transceiver architecture \cite{Xing18GC}. A pseudorandom noise (PN) sequence of length 2047 at 500 MHz was generated at baseband, then upconverted to a center frequency of 142 GHz, and transmitted through a directional and steerable horn antenna with 8\degree~half power beamwidth (HPBW) and 27 dBi gain from the TX. The signal was captured by the RX via an identical steerable horn antenna and downconverted to baseband signals. The baseband signal was sliding correlated with a local copy of the transmitted PN sequence at a slower rate \cite{Rap15textbook}, resulting power delay profile (PDP), which was sampled by a high-speed oscilloscope. TX and RX horn antennas were mechanically steered by two gimbals in the azimuth and elevation planes. The RX horn antenna was switched between vertical and horizontal polarization while the TX antenna remained vertically polarized. Both co- and cross-polarization measurements followed the identical measurement procedure as described below. The 142 GHz channel sounder had a null-to-null RF bandwidth of 1 GHz and can measure the directional propagation path loss up to 152 dB \cite{Xing18GC}.
\subsection{Measurement Environments and Locations}
\begin{figure}[h!]
	\vspace{-10pt}
		\setlength{\belowcaptionskip}{-15pt}
	\centering
	\includegraphics[width=1\linewidth]{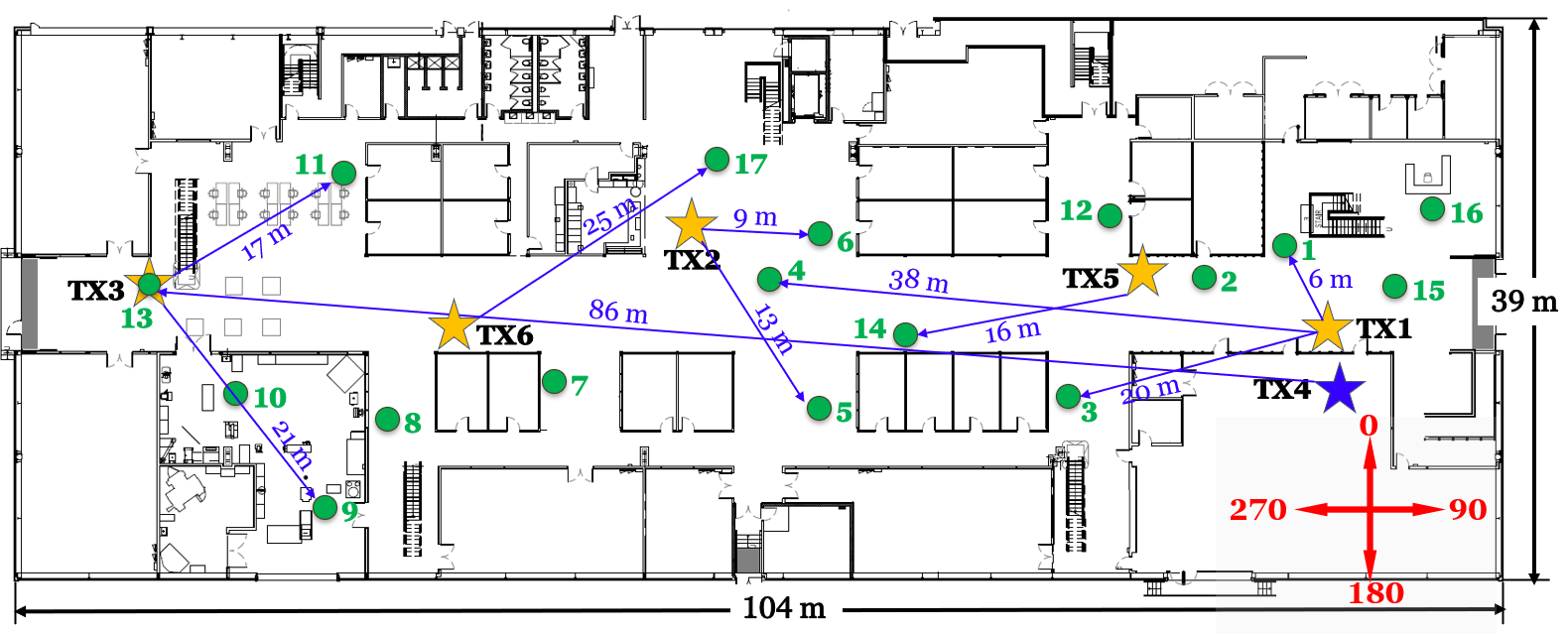}
	\caption{TX and RX locations in Factory A. Six TX locations are denoted as stars, and 17 RX locations are denoted as circles, resulting in 27 TX-RX location pairs for channel measurements. }
	\label{fig:floor_plan_newlab}
\end{figure}

The 142 GHz propagation measurements were conducted in four factory buildings in Brooklyn, NY. The selected factory buildings constitute a wide range of factory size, layout, and manufacturing facility. TX locations were picked to be typical of where WiFi access points are deployed to cover entire factory floors, and RX locations were selected along common trajectories of human mobile workers and AGVs in factories.   
\paragraph{Site A: Factory Building}The building ($\sim$104 m L$\times$ 39 m W $\times$ 25 m H) was partitioned into manufacturing labs and conference rooms covered by large glass windows with metal frames. A mezzanine floor was built above the rooms along external facades with a height of 3.7 m above the ground, leaving a 7 m wide open corridor at the center of the building. Six TX locations and 17 RX locations were selected to cover the entire open area in Factory A. Fig. \ref{fig:floor_plan_newlab} shows the 27 TX-RX location pairs consisting of 10 LOS locations, 15 NLOS locations, and two outage locations. RX12 and RX16 were in outage. The TX was set to 3 m above the ground, except that TX4 was set at 3 m high on the mezzanine floor, having a height of 6.7 m above the ground. The RX was set to 1.5 m above the ground. The T-R separation distance ranged from 6.5 m to 86.5 m. 

\begin{figure}[h!]
		\vspace{-10pt}
	\setlength{\belowcaptionskip}{-10pt}
	\centering
	\includegraphics[width=1\linewidth]{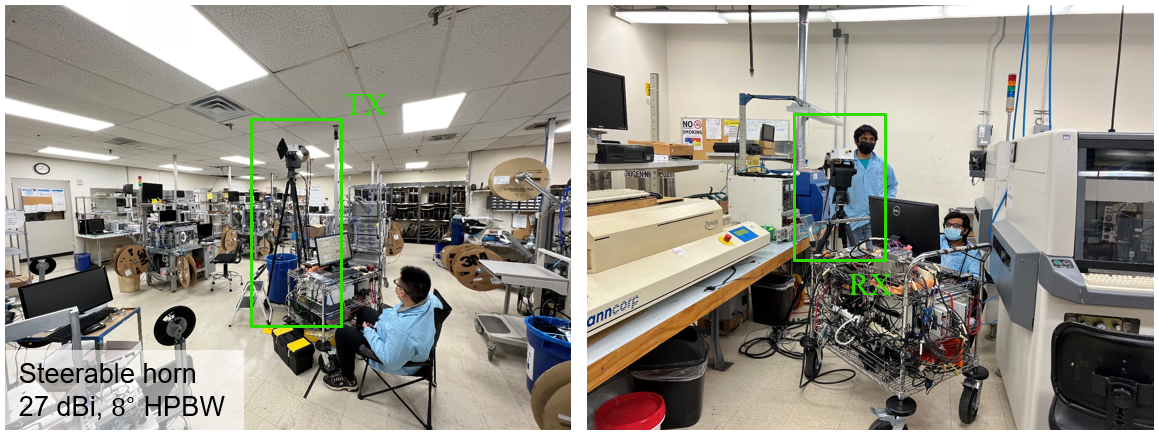}
	\caption{Photos taken from Factory B. The TX was set at 2.5 m, and the RX was set at 1.5 m.}
	\label{fig:photo_neptune}
\end{figure}

\paragraph{Site B: Electronics Manufacturing Facility} The single-story structure covered various production phases, such as manufacturing, testing, and assembly. We selected three production areas including automated manufacturing, testing and assembly for channel measurements. Factory B had heavy clutter containing massive metal objects such as robotics, partitions, machineries, and workstations as shown in Fig. \ref{fig:photo_neptune}. We selected four TX locations and 20 RX locations, covering the three areas of interest. The 20 TX-RX location pairs consisted of three LOS locations and 17 NLOS locations. The TX height was set to 2.5 m above the ground near the ceiling, and the RX height was set to 1.5 m above the ground. The T-R separation distance ranged from 5.6 m to 31.0 m. 

\begin{figure}[h!]
		\vspace{-10pt}
	\setlength{\belowcaptionskip}{-10pt}
	\centering
	\includegraphics[width=1\linewidth]{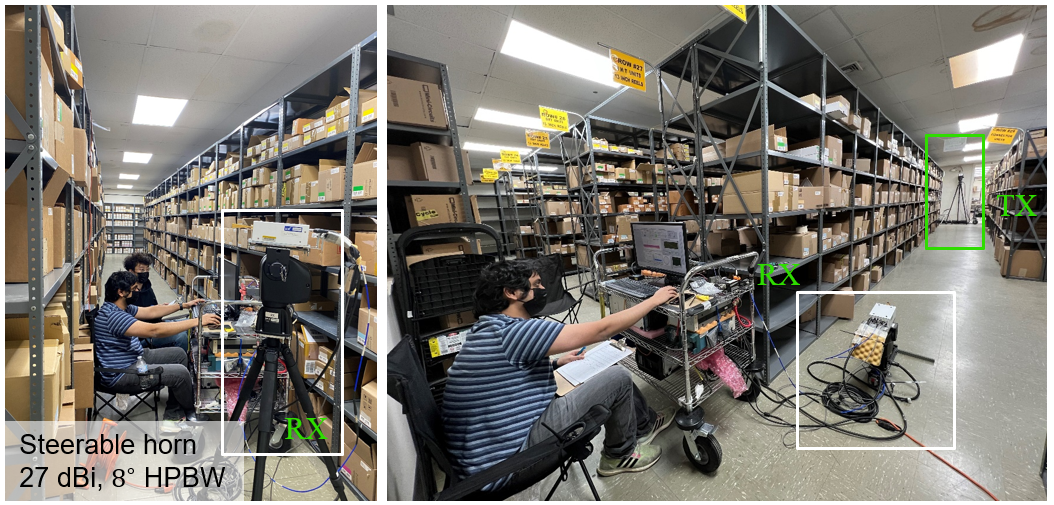}
	\caption{Photos taken from Factory C. The TX was set at 2.5 m, and the RX was set at 1.5 m and 0.5 m.}
	\label{fig:photo_deerpark}
\end{figure}

\paragraph{Site C: Warehouse Facility ($\sim$72 m L$\times$ 30 m W $\times$ 4 m H)}
Site C was a mid-size warehouse. As shown in Fig. \ref{fig:photo_deerpark}, large metal multi-layer shelves with a height of 3 m and a spacing distance of 1.5 m were placed at two sides of the building, creating a 2 m wide walkway in the center. These large metal shelves were filled with cardboard boxes containing commodities. We selected three TX locations and 11 RX locations in Factory C, covering the stock and shipping areas of the warehouse facility. The 11 TX-RX location pairs consisted of three LOS locations, six NLOS locations, and two outage locations. RX1 and RX8 were in outage. The TX was 2.5 m above the ground, and the RX was 1.5 m above the ground. The T-R separation distance ranged from 7.7 m to 37.6 m. In addition, the low RX height at 0.5 m above the ground was adopted at three LOS locations to study the effect of antenna height and ground reflection.
%\begin{figure}[h!]
%	\centering
%	\includegraphics[width=1\linewidth]{figs/floor_plan_deerpark.png}
%	\caption{TX and RX locations in Factory C. Three TX locations are denoted as stars, and 11 RX locations are denoted as circles, resulting in 11 TX-RX location pairs for channel measurements. }
%	\label{fig:floor_plan_deerpark}
%\end{figure}

\begin{figure}[h!]
		\vspace{-20pt}
	\setlength{\belowcaptionskip}{-10pt}
	\centering
	\includegraphics[width=1\linewidth]{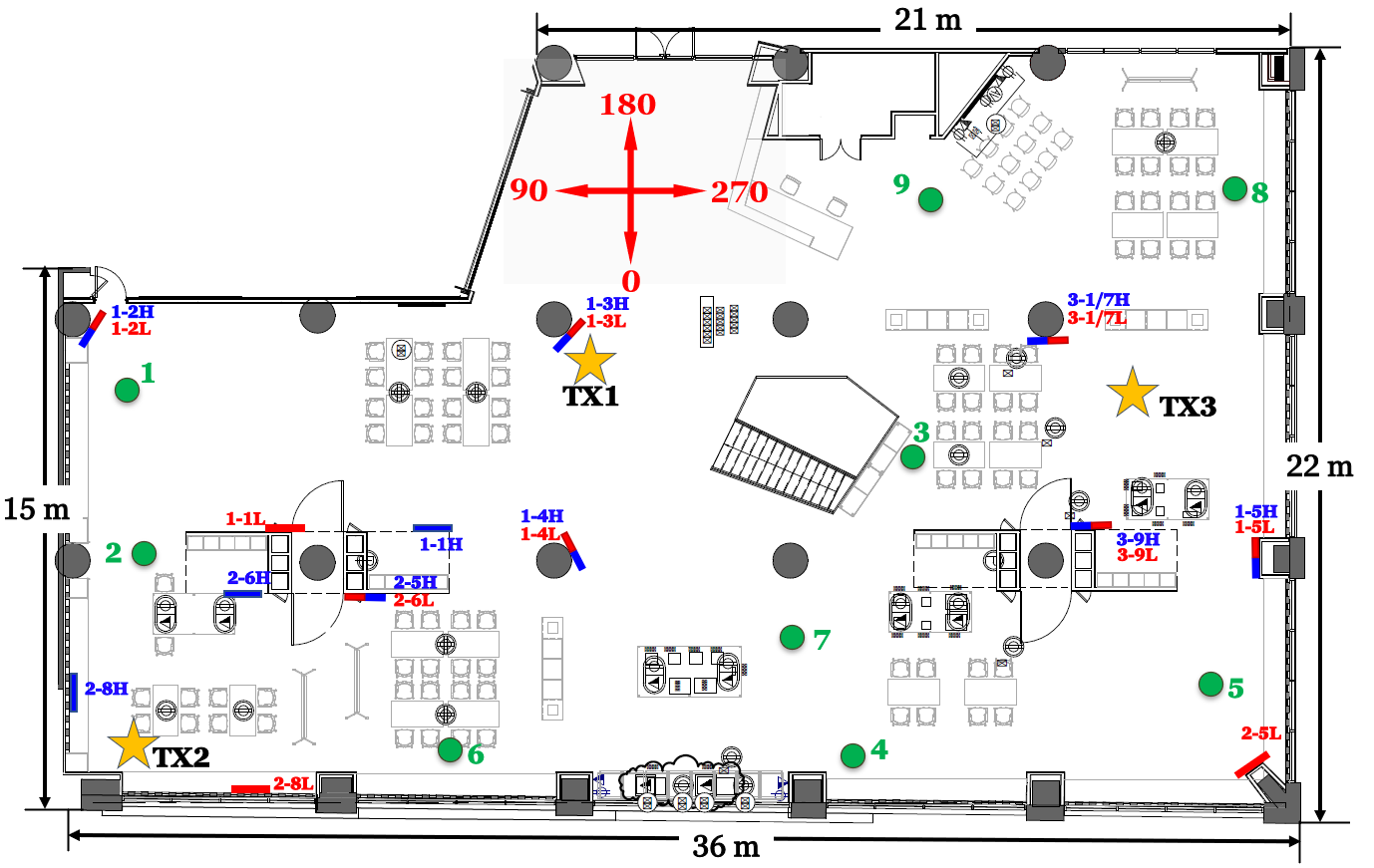}
	\caption{TX and RX locations in Factory Site D. Three TX locations are denoted as stars, and nine RX locations are denoted as circles, resulting in 12 TX-RX location pairs for channel measurements. The short blue and red bars denote the PRS positions for high and low RXs, respectively.}
	\label{fig:floor_plan_makerspace}
\end{figure}

\paragraph{Site D: Manufacturing Space ($\sim$35 m L$\times$ 25 m W $\times$ 5 m H)}
This space was a single large room for lightweight manufacturing and production, with various machineries such as laser cutters, CNC machines, and 3D printers. Large glass window panels coated with metallic frames covered the exterior walls on the north, south, and east of this room, whereas thick drywall constituted the wall on the west, as shown in Fig. \ref{fig:floor_plan_makerspace}. Plastic tables and wooden chairs were placed in a few office regions. Black-filled circles in Fig. \ref{fig:floor_plan_makerspace} represent concrete pillars which created strong reflections during channel measurements. 

We selected three TX locations and nine RX locations in Factory D, covering the entire manufacturing space. The TX was set to 2.5 m above the ground, and the RX was set to 1.5 m and 0.5 m above the ground at each RX location for studying the impact of antenna heights. The 12 TX-RX location pairs consisted of three LOS locations and nine NLOS locations for high RX. In contrast, only TX1 and RX3 were in LOS for low RX. The T-R separation distance ranged from 7.0 m to 35.1 m. In addition, the PRS measurements were conducted at twelve locations at both RX heights.

\subsection{Measurement Procedure}
The TX and RX were placed at the designated locations with vertical-to-vertical (V-V) antenna polarization. The TX and RX antenna pointing angles in the azimuth and elevation planes receiving the strongest power ($P_{\textup{max}}$) were identified as ``best'', which was the boresight direction for LOS locations or antennas pointing to a strong reflection for NLOS locations. The TX antenna was rotated from the best pointing angle in step increments of the antenna half power beamwidth (HPBW) (8\degree) in the azimuth plane, resulting in 45($=360\degree/8\degree$) TX pointing angles. With the TX pointed at the $i^{\textup{th}}$ pointing angle, the RX antenna performed a 360\degree~continuous sweep in the azimuth plane, taking about six seconds, and recorded the peak received power ($P^i_{\textup{peak}}$). We selected the TX angles based on a power threshold, which was set to the smaller value between 30 dB below the highest peak power from 45 TX angles, and 10 dB above the average noise floor. Those TX angles in which the peak received power exceeded the specified threshold were considered ``good'' angles, where detectable multipath existed. For each of the ``good'' TX pointing angles, the RX was rotated at step increments of 8\degree~in the complete azimuth plane. One directional PDP was recorded at each RX pointing step. The RX was then up- and down-tilted by 8\degree, and the exact stepped azimuthal sweep to acquire signal energy was repeated \cite{Sun15GC}. For example, if a TX has ten ``good'' pointing angles, there are at most $10\times45\times3=1350$ possible directional PDPs measured at this location. In practice, we measured 1000 PDPs on average at each particular TX-RX location. This measurement procedure is exhaustive, taking typically two hours for each TX-RX location, but provides complete angular statistics for both angle of departure (AOD) and angle of arrival (AOA). Identical measurements were performed for both V-V and V-H antenna polarization configurations at each TX-RX location pair. 

\section{142 GHz Path Loss Models for Co- and Cross-Polarization Antennas}
\label{sec:pathloss}
We adopt the well-known close-in free space reference distance (CI) path model for omnidirectional and directional path loss measurement data because the CI path loss model has been proven to be superior for modeling path loss over many environments and frequencies \cite{Sun16TVT}. The CI path loss model is parameterized solely by the path loss exponent (PLE) denoted as $n$. $\PL^\textup{CI}$ represents the path loss in dB scale, which is a function of distance and frequency \cite{Rap15TCOMM,Ju21JSAC,Xing21CLb}:
\begin{align}\label{eq:CI1}
	\PL(f, d)[\dB]=&\textup{FSPL}(f, d_0) +10n\log_{10}\left(\frac{d}{d_0}\right)+\chi_{\sigma}^{\mathrm{CI}} \text{,}
\end{align}
for $d\geq d_0$, where $d_0 = 1 m$. $d$ is the 3-D T-R separation distance and $d_0$ is the reference distance. The free space path loss at a reference distance of $d_0=1$ m at the carrier frequency $f$ is given by $\mathrm{FSPL}(f, 1~\textup{m})[\dB]=20\log_{10}\left(4\pi f\times 10^9/c\right)=32.4[\dB]+20\log_{10}(f)$. $\chi^\mathrm{CI}_{\sigma}$ denotes shadow fading (SF), commonly modeled as a normal random variable with zero mean and $\sigma$ standard deviation in dB. The CI path loss model uses the FSPL at the reference distance as a fixed intercept and finds a straight line fit to the measured path loss data that achieves the minimum mean square error \cite{Rap15TCOMM, Ju21JSAC, Xing21CLb}.

\begin{figure}[h!]
	\vspace{-10pt}
	\setlength{\belowcaptionskip}{-10pt}
	\centering
	\includegraphics[width=1\linewidth]{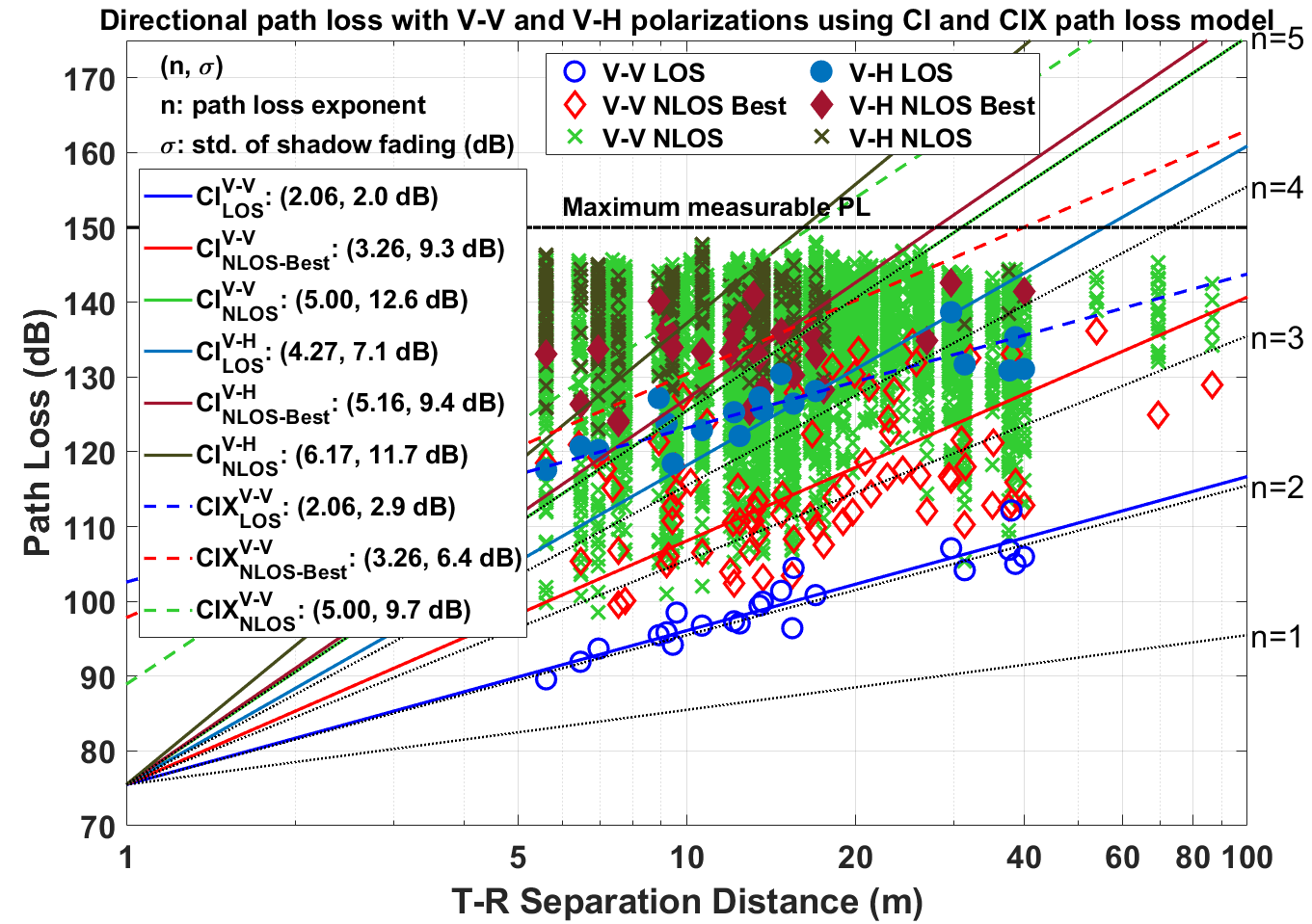}
	\caption{Directional path loss data and best fit CI path loss models ($d_0=1$ m) for V-V polarizations for 142 GHz indoor factory channels from four factories.}
	\label{fig:dir_pathloss_all}
\end{figure}
Fig. \ref{fig:dir_pathloss_all} shows the directional path loss scatter plots and best fit CI (\ref{eq:CI1}) models for the 142 GHz LOS and NLOS scenarios for the co-polarized (V-V) and cross-polarized (V-H) antenna configurations from all four factory sites. Each scatter point denotes a unique pointing direction pair of TX and RX horn antennas with 8\degree~HPBW. The LOS direction denotes the unique TX-RX angles when the TX and RX point to each other. The NLOS direction denotes all the remaining TX-RX angles in which the receive power exceeds the threshold. The Best NLOS direction is picked from the NLOS directions for each TX-RX location pair, which contains the most received power. Note that the directions adjacent to the LOS direction are excluded from the selection of NLOS Best directions to avoid the effect of antenna pattern. 

For V-V antenna polarization, the LOS PLE is 2.06, close to the free space PLE of 2. The NLOS PLE is reduced from 5.01 to 3.27 by pointing antennas to the strongest NLOS direction instead of an arbitrary direction, suggesting that beam steering is critical for THz communication systems. However, Site A and Site D produce smaller NLOS and NLOS-Best PLEs than Site B and Site C, possibly because Site A and Site D are more open-planned with fewer partitions and less clutter density. For the V-H polarized measurements, the LOS, NLOS-Best, and NLOS PLEs are 4.27, 5.16, and 6.17, respectively, which are much greater than the PLEs for V-V antenna polarization due to the polarization mismatch between antennas and channels. Using the PLEs of co-polarized path loss data, the CIX model \cite{Ju22ICC} provides a much better fit than the CI model by reducing the standard deviation from 7.1 dB to 2.9 dB for LOS directions and from 9.4 dB to 6.4 dB for NLOS-Best directions. The XPD for LOS directions is 27.1 dB, close to the antenna XPD measured in free space of 27.5 dB \cite{Shakya21TCS}. The NLOS-Best directions produce an XPD of 22.2 dB, and the NLOS directions produce an XPD of 13.3 dB. The NLOS-Best directions show an XPD reduction of 5 dB compared to the LOS direction, indicating that reflection and scattering have a prominent de-polarization effect. 

\begin{figure}[h!]
		\vspace{-10pt}
	\setlength{\belowcaptionskip}{-10pt}
	\centering
	\includegraphics[width=1\linewidth]{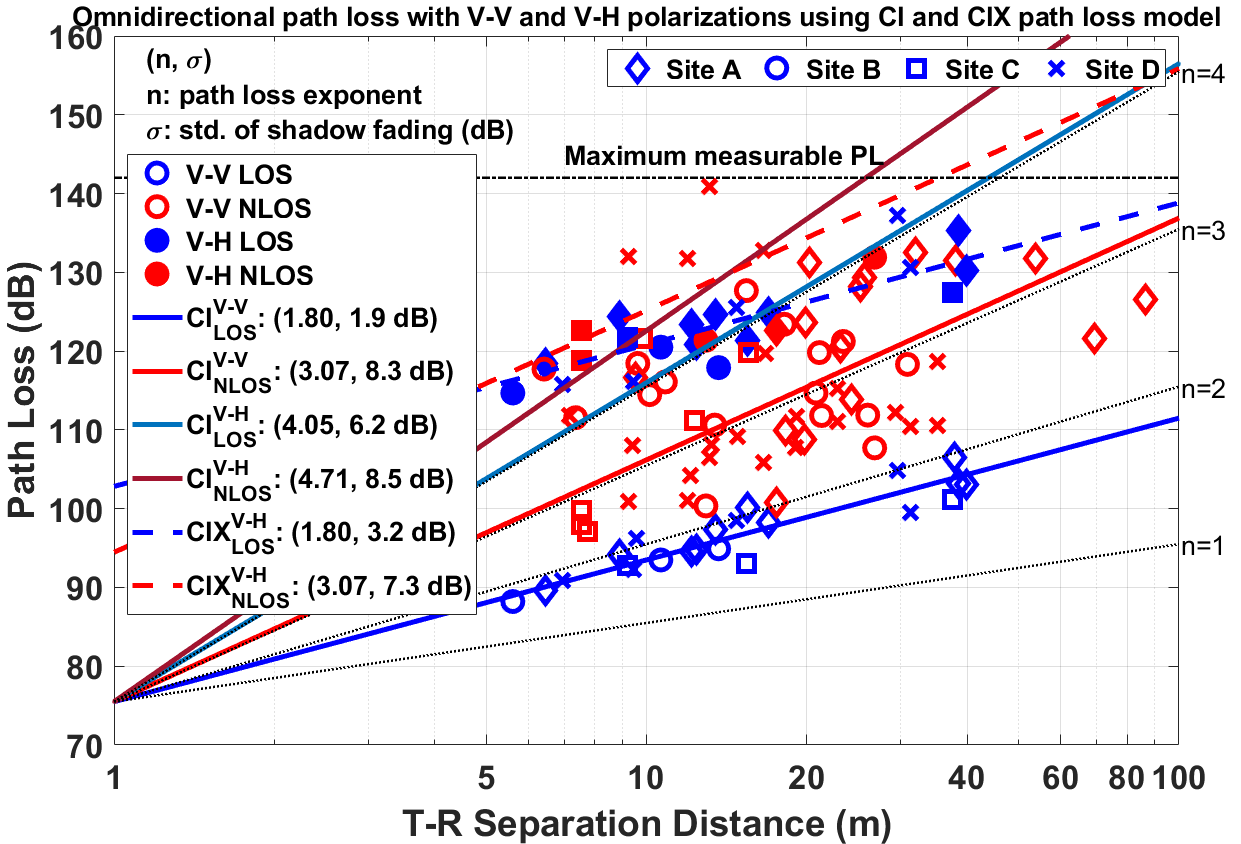}
	\caption{Omnidirectional CI and CIX path loss model parameters with $d_0=1$ m for 142 GHz indoor factory channels for V-V and V-H antenna polarization orientations in four factories. }
	\label{fig:omni_pathloss_all}
\end{figure}

Fig. \ref{fig:omni_pathloss_all} displays the omnidirectional path loss scatter plot and best fit CI and CIX models at 142 GHz in LOS and NLOS environments for co-polarization (V-V) and cross-polarization (V-H) antenna configurations from all four factory sites. Factory sites are denoted as different markers. For V-V polarization, the LOS PLE is 1.80, which is similar to the value in office environments (1.75) \cite{Ju21JSAC,Xing21CLa} and better than the outdoor environments (1.91) \cite{Xing21ICC}. The NLOS PLE is 3.07, which is worse than 2.7 in office environment \cite{Ju21JSAC,Xing21CLa} and 2.9 in outdoor environment \cite{Ju21GC,Xing21CLb}, indicating that the factories may be harsh for signal propagation in NLOS scenarios due to 1) dense clutter such as workstations and shelves, 2) large space leading to long propagation distances and a high chance of blockage, 3) smooth surfaces of metallic objects producing only specular reflections with limited scattering. Previous works on channel measurements in factories at mmWave frequencies showed similar PLEs \cite{Schmieder19GC,Saman21Sensor,Chizhik19ISAP}.

\section{High and low RX heights}
\label{sec:rx_height}
According to the terminologies from 3GPP TR 38.901 \cite{3GPP38901r16}, a device is clutter-elevated if the antenna is higher than the average environmental clutter height. A device is clutter-embedded if the device antenna is lower than the average environmental clutter height. We conducted comparative propagation measurements at 1.5 m and 0.5 m RX heights emulating high (clutter-elevated) UEs and low (clutter-embedded) UEs at factory site D, where the average clutter height is 1 m. The TX was fixed at 2.5 m as indoor access points. As shown in Fig. \ref{fig:floor_plan_makerspace}, three TX locations and nine RX locations were selected in Site D, resulting in 12 TX-RX location pairs. Both high and low UEs were measured at 12 identical locations. For the high UE measurements, there were five LOS location pairs and seven NLOS location pairs. There were one LOS location pair and eleven NLOS location pairs for the low UE measurements. 
\begin{figure}[h]
	\setlength{\belowcaptionskip}{-10pt}
	\centering
	\includegraphics[width=.9\linewidth]{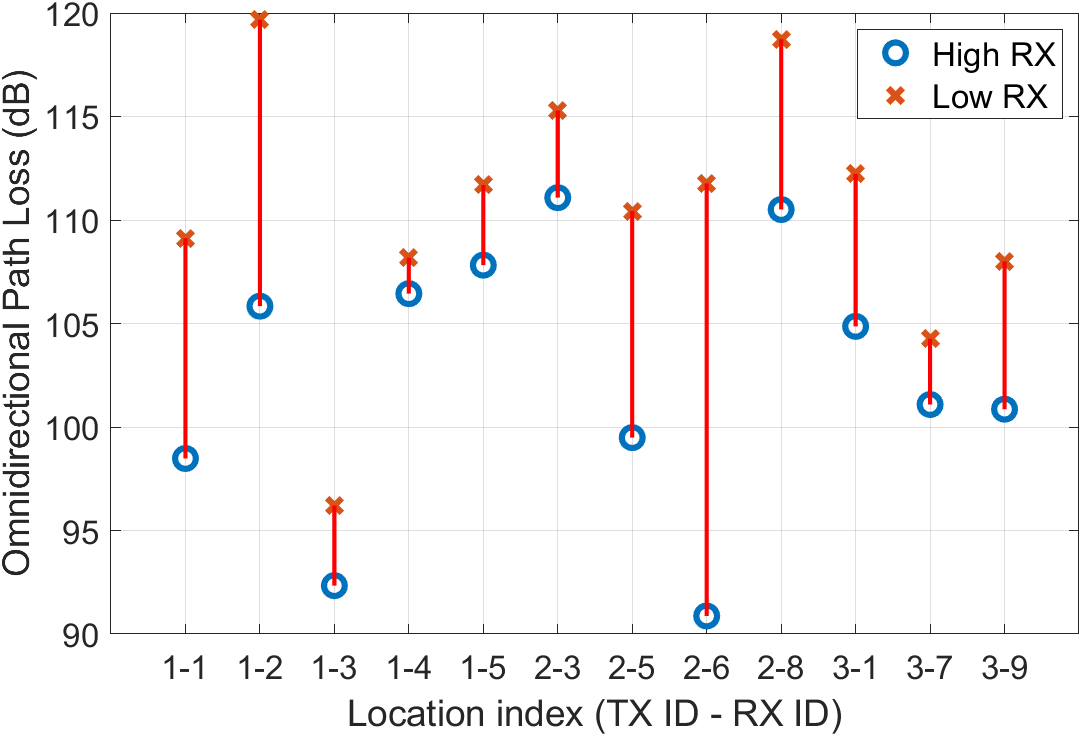}
	\caption{Location-wise comparison of omnidirectional path loss for high and low RXs in factory site D.}
	\label{fig:omni_pl_change_ht}
\end{figure}

Fig. \ref{fig:omni_pl_change_ht} compares omnidirectional path loss between the high and low RX heights at each measurement location at Factory Site D. The omnidirectional path loss increases at every LOS and NLOS location because the minimum path loss variation is -1.7 dB (i.e., the high-UE path loss is smaller than the low-UE path loss). The maximum path loss increases are 20.8 dB and 13.8 dB in the LOS and NLOS scenarios. The mean path loss increases are 10.7 dB and 6.0 dB in the LOS and NLOS scenarios. The larger path loss increase in the LOS scenario is expected due to the potential blockage of the LOS path. For example, the largest path loss increase was recorded at TX2 and RX6, where a 1 m tall machine blocked the boresight path between TX2 and RX6. The larger standard deviation of 6.3 dB in the LOS scenario also suggests the risk of the diminishing LOS path. The derived NLOS PLEs are 2.50 and 2.88 for the high and low RXs, respectively, indicating that the low RX often experiences additional path loss due to the blockage and shadowing effect of clutter in the proximity.

\section{Channel Enhancement Using Passive Reflecting Surfaces}
\label{sec:surface}
We conducted regular channel measurements at each location without the PRS by rotating the TX and RX antennas in the azimuth and elevation planes to record omnidirectional channel responses. The PRS position was selected based on the regular measurement results to improve a single-bounce reflection path. Next, a visual ray tracing was performed to identify the reflecting object. Only the reflecting objects such as walls, pillars, corners, and large fixed furniture were considered candidate PRS locations because hard or flexible RISs would be installed on fixed large surfaces in future indoor scenarios. The metal plate was placed next to the reflecting object at a height between the TX and RX, depending on the relative positions. We manually and carefully rotated the metal plate in the horizontal plane to find the best orientation that provided the largest channel gain, which emulated the configurability of RISs steering the beam to the best direction based on channel estimation. With the TX antenna fixed, the RX antenna was swept over 360\degree~in the azimuth plane in step increments of the antenna half-power beamwidth (HPBW) of 8\degree. Identical azimuth sweeps were performed at three elevation planes, which were boresight, 8\degree~above and below the boresight. The adjacent azimuth TX pointing directions with 8\degree~spacing that showed sufficient reflecting power were measured at several locations. 
%\begin{figure}
%	\centering
%	\subfloat[Directional path loss reduction with PRS aid in factory site D.]{
%		\label{fig:dir_pl_irs}
%		\includegraphics[width=.9\linewidth]{figs/dir_pl_irs.png}}
%	\hfill
%	\subfloat[Omnidirectional path loss reduction with PRS aid in factory site D.]{
%		\label{fig:omni_pl_irs}
%		\includegraphics[width=.9\linewidth]{figs/omni_pl_irs.png} } 
%	\caption{Directional and omnidirectional path loss comparison between the natural and the PRS-aided channels measured at high (H) and low (L) RX heights in factory site D. }
%	\label{fig:both_pl_irs}
%	\vspace{-10pt}
%\end{figure}

\begin{figure}
	\centering
	\begin{subfigure}[b]{.9\linewidth}
		\centering
		\includegraphics[width=\textwidth]{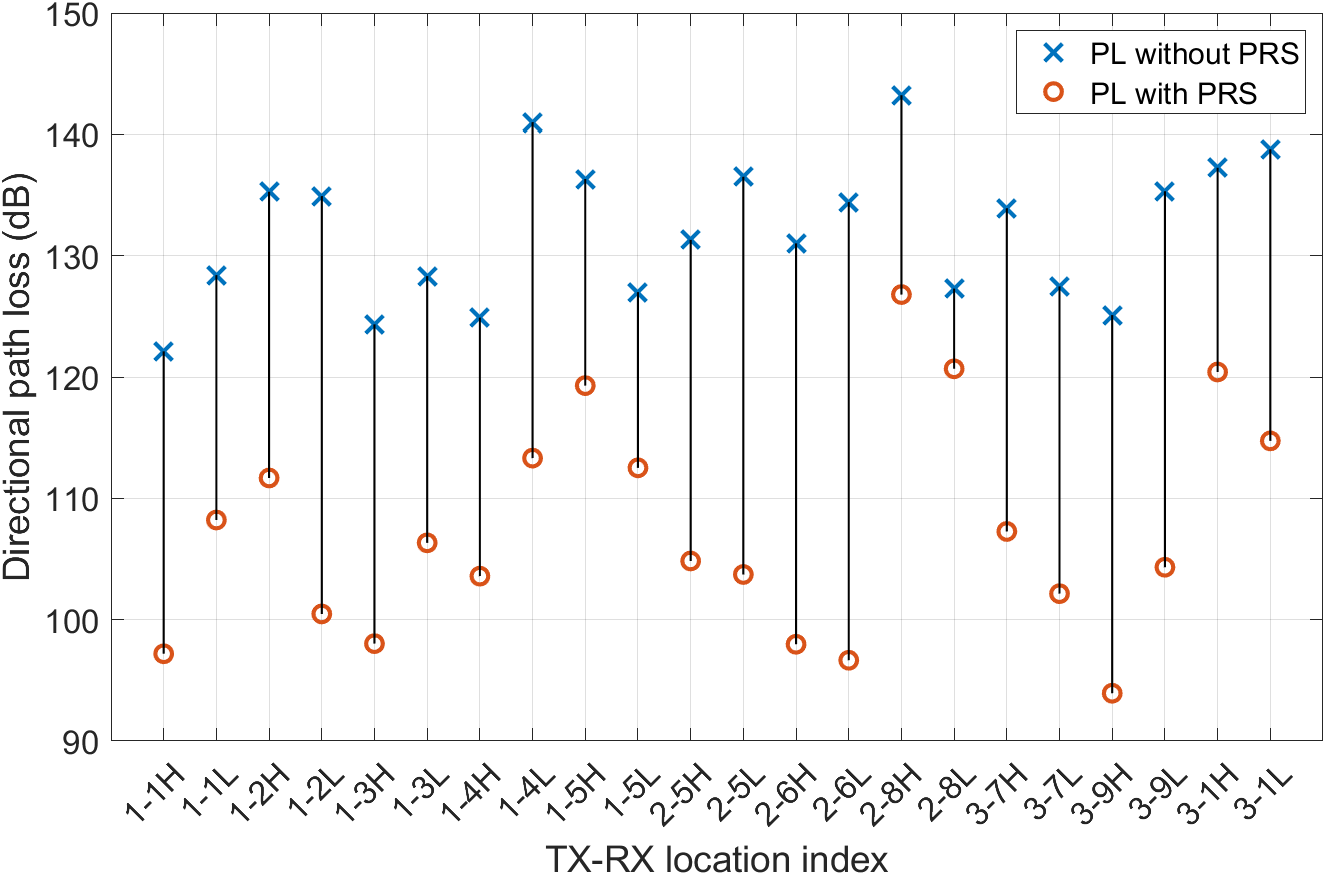}
		\caption{Directional path loss reduction with PRS aid at high (H) and low (L) RX heights in factory site D.}
		\label{fig:dir_pl_irs}
	\end{subfigure}
	\hfill
	\begin{subfigure}[b]{.9\linewidth}
		\centering
		\includegraphics[width=\textwidth]{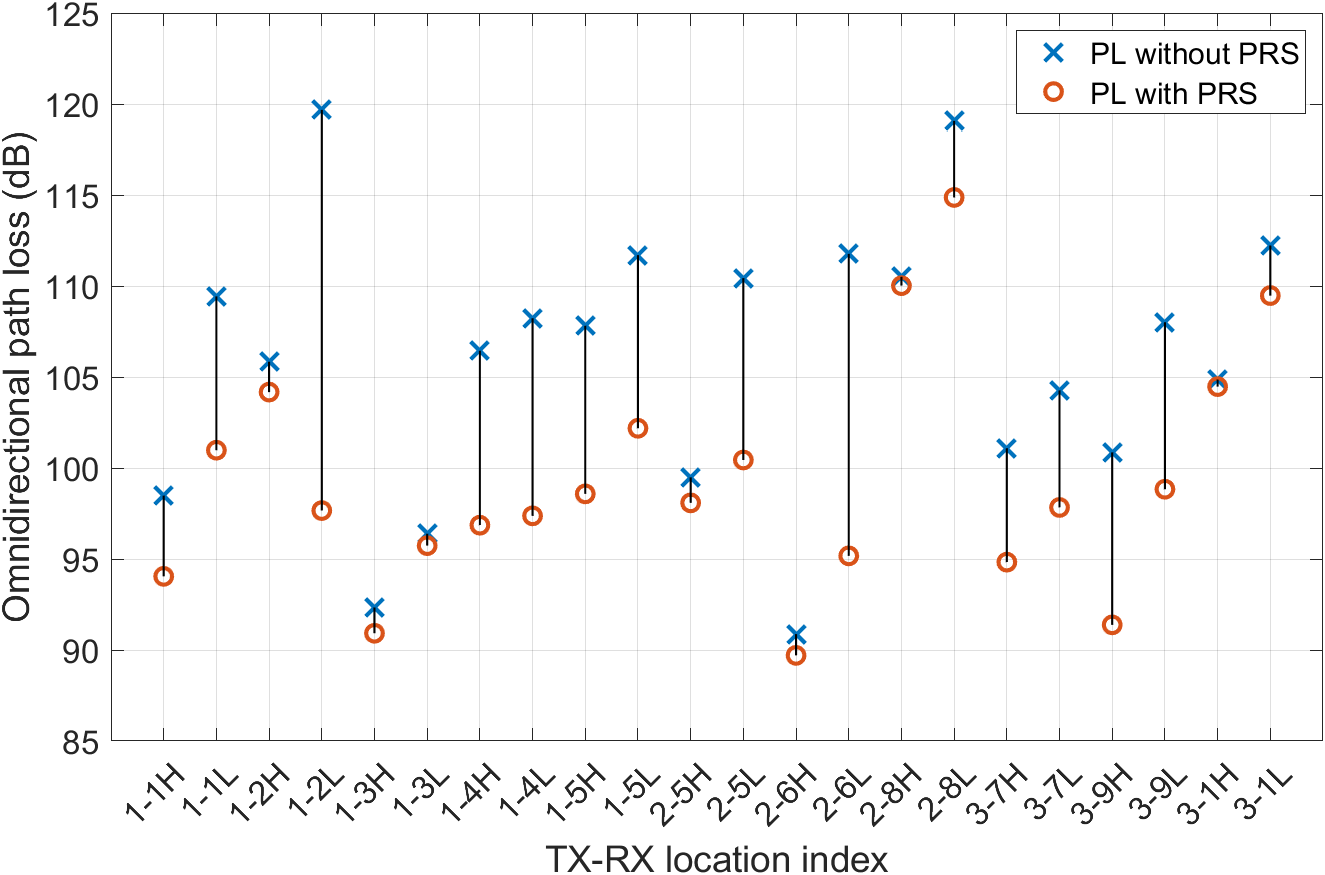}
		\caption{Omnidirectional path loss reduction with PRS aid at high (H) and low (L) RX heights in factory site D.}
		\label{fig:omni_pl_irs}
	\end{subfigure}
	\caption{Directional and omnidirectional path loss comparison between the natural and the PRS-aided channels measured at high (H) and low (L) RX heights in factory site D. }
	\label{fig:both_pl_irs}
	\vspace{-10pt}
\end{figure}
The directional path loss reduction in the reflecting direction enhanced by the PRS is shown in Fig. \ref{fig:dir_pl_irs} location by location, where, for each TX-RX location pair, the TX-RX antenna pointing combination that presents the maximum directional power gain is plotted. Note that the directional path loss power is calculated by summing all the samples recorded in a directional PDP, which may contain multiple MPCs. The PRS provides a mean power increase of 25 dB with a maximum of 37.8 dB and a minimum of 6.6 dB. The TX2 and low RX6 show the maximum gain of 37.8 dB because the distance between the TX2, PRS2-6L, and RX6 is the shortest among all the measurement locations. In addition, the reflection from the partition wall where PRS2-6L is located is partially blocked by the machinery near the RX6 without the aid of a PRS. However, the PRS emulates the anomalous reflector that does not obey the specular reflection from the original furniture surface and creates a strong single-bounce reflection path. The TX2 and low RX8 present the minimum directional path power gain of 6.6 dB because the signal is transmitted over the largest T-R separation distance of 38 m measured in site D, where many machines and tables are located between the TX and RX causing partial blockage. Thus, the link created by the PRS does not have a clear first-order reflection path, indicating that the choice of location of PRS2-8L might not be optimal. Fig. \ref{fig:omni_pl_irs} shows the influence of the PRS on omnidirectional path loss. The PRS increases the omnidirectional received power by a maximum of 22 dB and a minimum of 0.5 dB with a mean of 6.5 dB. This increase depends on the PRS and other existing MPCs. A slight increase indicates that other MPCs dominate the total received power, which is typical for the LOS scenario. However, a significant increase often occurs at NLOS and near-outage locations, where the PRS-enhanced path becomes the strongest path and dominates the channel. 

\section{Conclusions}
\label{sec:conclusion}
This paper presents the first extensive sub-THz channel measurement campaign at 142 GHz for the indoor factory scenarios. The directional and omnidirectional path loss are modeled based on 82 TX-RX locations, including 30 LOS locations, 48 NLOS locations, and four outage locations. The omnidirectional PLE for NLOS locations is 3.1, which is higher than the values obtained from indoor office (2.7) and outdoor urban environments (2.9), suggesting that factories may induce additional attenuation due to heavy clutter for NLOS environments. Furthermore, the channel measurements conducted at a low RX height indicate that near-the-ground UEs such as AGVs may experience larger path loss (i.e., 10.7 dB and 6.0 dB increase for LOS and NLOS locations) compared to a high RX height due to a higher probability of blockage and scattering loss caused by surrounding obstructions. Moreover, the channel enhancement measurements demonstrate that a single large metal plate can provide an average gain of 25 dB in the desired pointing directions by accurately orientating the plate. In general, the PRS increases the omnidirectional received power by a maximum of 22 dB and a minimum of 0.5 dB with a mean of 6.5 dB, suggesting that future implementations and deployments of RISs can play an essential part in sub-THz wireless systems. This work may motivate more effort on sub-THz applications in factories and provide insights on factory channel model standardization above 100 GHz.  
\bibliographystyle{IEEEtran}
\bibliography{icc23}

\end{document}